\pdfoutput=1
\documentclass[12pt,a4paper]{article}

\usepackage{ifthen} 
\usepackage{verbatim}
\usepackage{overpic}

\newboolean{pdflatex}
\setboolean{pdflatex}{true} 

\newboolean{articletitles}
\setboolean{articletitles}{true} 

\newboolean{uprightparticles}
\setboolean{uprightparticles}{false} 

\def\paperauthors{Chenxu~Yu, Yanxi~Zhang} 
\def\paperasciititle{ABC} 
\def\papertitle{A method of maximum likelihood fit to data with non-uniform efficiencies} 
\def\paperkeywords{{High Energy Physics}, {Fit}}

\def\paperlicenceurl{https://creativecommons.org/licenses/by/4.0/}

\usepackage[top=1in, bottom=1.25in, left=1in, right=1in]{geometry}

\usepackage{microtype}
\usepackage{lineno}  
\usepackage{xspace} 
\usepackage{caption} 

\usepackage{graphicx}  
\usepackage{color}
\usepackage{colortbl}
\graphicspath{{./figs/}} 

\usepackage{amsmath} 
\usepackage{amssymb}
\usepackage{amsfonts}
\usepackage{upgreek} 

\newcommand*\patchAmsMathEnvironmentForLineno[1]{%
\expandafter\let\csname old#1\expandafter\endcsname\csname #1\endcsname
\expandafter\let\csname oldend#1\expandafter\endcsname\csname
end#1\endcsname
 \renewenvironment{#1}%
   {\linenomath\csname old#1\endcsname}%
   {\csname oldend#1\endcsname\endlinenomath}%
}
\newcommand*\patchBothAmsMathEnvironmentsForLineno[1]{%
  \patchAmsMathEnvironmentForLineno{#1}%
  \patchAmsMathEnvironmentForLineno{#1*}%
}
\AtBeginDocument{%
\patchBothAmsMathEnvironmentsForLineno{equation}%
\patchBothAmsMathEnvironmentsForLineno{align}%
\patchBothAmsMathEnvironmentsForLineno{flalign}%
\patchBothAmsMathEnvironmentsForLineno{alignat}%
\patchBothAmsMathEnvironmentsForLineno{gather}%
\patchBothAmsMathEnvironmentsForLineno{multline}%
\patchBothAmsMathEnvironmentsForLineno{eqnarray}%
}

\usepackage{hyperxmp}

\usepackage[pdftex,
            pdfauthor={\paperauthors},
            pdftitle={\paperasciititle},
            pdfkeywords={\paperkeywords},
            pdflicenseurl={\paperlicenceurl}]{hyperref}

\usepackage[colorinlistoftodos,textsize=scriptsize]{todonotes}

\usepackage[bottom,flushmargin,hang,multiple]{footmisc}

\usepackage[all]{hypcap} 

\usepackage{xspace} 
\usepackage{upgreek}

\def\MagUp {\mbox{\em Mag\kern -0.05em Up}\xspace}

\ifthenelse{\boolean{uprightparticles}}%
{

 \def\PDelta      {\ensuremath{\Delta}\xspace}                 
 \def\PXi         {\ensuremath{\Xi}\xspace}                 
 \def\PLambda     {\ensuremath{\Lambda}\xspace}                 
 \def\PSigma      {\ensuremath{\Sigma}\xspace}                 
 \def\POmega      {\ensuremath{\Omega}\xspace}                 
 \def\PUpsilon    {\ensuremath{\Upsilon}\xspace}
 \let\oldPi\Pi
 \def\PPi         {\ensuremath{\oldPi}\xspace}

 \def\PB      {\ensuremath{\mathrm{B}}\xspace}                 
                  
 \def\PD      {\ensuremath{\mathrm{D}}\xspace}

 \def\PK      {\ensuremath{\mathrm{K}}\xspace}

 \def\Pi      {\ensuremath{\mathrm{i}}\xspace}

 \def\Ps      {\ensuremath{\mathrm{s}}\xspace}

 \def\thebaroffset{0.0em}
}
{

 \mathchardef\PDelta="7101
 \mathchardef\PXi="7104
 \mathchardef\PLambda="7103
 \mathchardef\PSigma="7106
 \mathchardef\POmega="710A
 \mathchardef\PUpsilon="7107
 \mathchardef\PPi="7105
                  
 \def\PB      {\ensuremath{B}\xspace}                 
                  
 \def\PD      {\ensuremath{D}\xspace}

 \def\PK      {\ensuremath{K}\xspace}

 \def\Pi      {\ensuremath{i}\xspace}

 \def\Ps      {\ensuremath{s}\xspace}

 \def\thebaroffset{0.18em}
}
\newcommand{\offsetoverline}[2][\thebaroffset]{\kern #1\overline{\kern -#1 #2}}%

\makeatletter
\ifcase \@ptsize \relax
  \newcommand{\miniscule}{\@setfontsize\miniscule{4}{5}}
\or
  \newcommand{\miniscule}{\@setfontsize\miniscule{5}{6}}
\or
  \newcommand{\miniscule}{\@setfontsize\miniscule{5}{6}}
\fi
\makeatother

\DeclareRobustCommand{\optbar}[1]{\shortstack{{\miniscule (\rule[.5ex]{1.25em}{.18mm})}
  \\ [-.7ex] $#1$}}

\def\squark    {{\ensuremath{\Ps}}\xspace}

\def\KorKbar {\kern \thebaroffset\optbar{\kern -\thebaroffset \PK}{}\xspace}

\def\D       {{\ensuremath{\PD}}\xspace}

\def\DorDbar {\kern \thebaroffset\optbar{\kern -\thebaroffset \PD}\xspace}

\def\Dp      {{\ensuremath{\D^+}}\xspace}
\def\Dm      {{\ensuremath{\D^-}}\xspace}

\def\DpDm    {\ensuremath{\Dp {\kern -0.16em \Dm}}\xspace}

\def\B       {{\ensuremath{\PB}}\xspace}

\def\BorBbar {\kern \thebaroffset\optbar{\kern -\thebaroffset \PB}\xspace}

\def\Bd      {{\ensuremath{\B^0}}\xspace}

\def\BdorBdbar {\kern \thebaroffset\optbar{\kern -\thebaroffset \Bd}\xspace}

\def\Bs      {{\ensuremath{\B^0_\squark}}\xspace}

\def\BsorBsbar {\kern \thebaroffset\optbar{\kern -\thebaroffset \Bs}\xspace}

\def\Y#1S{\ensuremath{\PUpsilon{(#1S)}}\xspace}

\def\LorLbar     {\kern \thebaroffset\optbar{\kern -\thebaroffset \PLambda}\xspace}

\def\to                 {\ensuremath{\rightarrow}\xspace}

\def\AT#1     {\ensuremath{A_{\mathrm{T}}^{#1}}\xspace}           

\def\C#1      {\ensuremath{\mathcal{C}_{#1}}\xspace}                       
\def\Cp#1     {\ensuremath{\mathcal{C}_{#1}^{'}}\xspace}                    
\def\Ceff#1   {\ensuremath{\mathcal{C}_{#1}^{\mathrm{(eff)}}}\xspace}        
\def\Cpeff#1  {\ensuremath{\mathcal{C}_{#1}^{'\mathrm{(eff)}}}\xspace}       
\def\Ope#1    {\ensuremath{\mathcal{O}_{#1}}\xspace}                       
\def\Opep#1   {\ensuremath{\mathcal{O}_{#1}^{'}}\xspace}                    

\newcommand{\aunit}[1]{\ensuremath{\text{\,#1}}}       

\newcommand{\tev}{\aunit{Te\kern -0.1em V}\xspace}
\newcommand{\gev}{\aunit{Ge\kern -0.1em V}\xspace}
\newcommand{\mev}{\aunit{Me\kern -0.1em V}\xspace}
\newcommand{\kev}{\aunit{ke\kern -0.1em V}\xspace}
\newcommand{\ev}{\aunit{e\kern -0.1em V}\xspace}
 
\newcommand{\mevc}{\ensuremath{\aunit{Me\kern -0.1em V\!/}c}\xspace}
\newcommand{\gevc}{\ensuremath{\aunit{Ge\kern -0.1em V\!/}c}\xspace}
\newcommand{\mevcc}{\ensuremath{\aunit{Me\kern -0.1em V\!/}c^2}\xspace}
\newcommand{\gevcc}{\ensuremath{\aunit{Ge\kern -0.1em V\!/}c^2}\xspace}

\def\gsim{{~\raise.15em\hbox{$>$}\kern-.85em
          \lower.35em\hbox{$\sim$}~}\xspace}
\def\lsim{{~\raise.15em\hbox{$<$}\kern-.85em
          \lower.35em\hbox{$\sim$}~}\xspace}

\def\tell1  {TELL1\xspace}
\def\ukl1   {UKL1\xspace}

\newcommand{\lhcborcid}[1]{\href{https://orcid.org/#1}{\hspace*{0.1em}\raisebox{-0.45ex}{\includegraphics[width=1em]{figs/orcidIcon.pdf}}}}

\usepackage{cite} 
\usepackage{mciteplus}

\usepackage{longtable}
\usepackage{multirow}
\usepackage{rotating}
\begin{document}

\renewcommand{\thefootnote}{\fnsymbol{footnote}}
\setcounter{footnote}{1}


\begin{titlepage}
\pagenumbering{roman}


\noindent

\vspace*{1.0cm}

{\normalfont\bfseries\boldmath\huge
\begin{center}
  \papertitle 

\end{center}
}

\vspace*{1.0cm}

\begin{center}
Chenxu Yu\footnote{2301110142@pku.edu.cn},
Yanxi Zhang\footnote{yanxi.zhang@pku.edu.cn}
  \bigskip\\
{\normalfont\itshape\footnotesize

Peking University, Beijing, China \\

}
\end{center}

\vspace{\fill}

\begin{abstract}
Estimations of physical parameters using data usually involve non-uniform experimental efficiencies. In this article, a method of maximum likelihood fit is introduced using the efficiency as a weight, while the probability distribution function is kept unaffected by the efficiency. A brief proof and pseudo-experiment studies suggest that this method gives unbiased estimation of parameters. For cases where the probability distribution function can be normalized analytically, this method significant reduces the usage of computing resources.

\end{abstract}

\vspace*{2.0cm}

\begin{center}

\end{center}

\vspace{\fill}


\end{titlepage}




\renewcommand{\thefootnote}{\arabic{footnote}}
\setcounter{footnote}{0}

\cleardoublepage


\pagestyle{plain}
\setcounter{page}{1}
\pagenumbering{arabic}

\section{Introduction}
The unbinned maximum likelihood fit serves as an indispensable tool for parameter estimation in the realm of high-energy physics, owing to the highly desirable characteristics of the maximum likelihood estimator. 
In the asymptotic limit, this estimator conforms to a normal distribution centered around the true parameter value, with a variance equivalent to the minimum variance bound. 
Furthermore, the unbinned approach ensures that no information is lost due to binning.

The incorporation of weights into the maximum likelihood formalism is advantageous in numerous applications. 
Data, from which unknown physical parameters in a predefined probability distribution function (PDF) are extracted, are usually folded by an experimental efficiency distribution. A non-uniform efficiency distribution  must be corrected to avoid biases in the estimation of physical parameters.
This article delves into the application of the inverse of per-event efficiency as the weight to the maximum likelihood to address the effect of experimental efficiency. 
The method is compared with the traditional one which uses the product of the parameter-encoded ideal PDF and the efficiency distribution in the maximum likelihood fit. 
The consistency between the two methods is thoroughly examined. 
A weighted maximum likelihood fit is known to result in a biased parameter variance~\cite{Langenbruch:2019nwe}. An additional global weight is introduced  to rectify the bias. In the following, a brief introduction and proof of the efficiency weighted maximum likelihood fit is provided, followed by an example to numerically validity the method.

\section{Unbined maximum likelihood construction}
Typically, the likelihood estimator for a set of $N_P$ parameters $\boldsymbol{\theta}=\{\theta_1,...,\theta_{N_P}\}$, given N independent measurements $S_e=\{\vec{x}_1,...,\vec{x}_N\}$ sampled according to the PDF $f(\vec{x},\boldsymbol{\theta}')$, is defined as

\begin{eqnarray}
    \ln \left[\mathcal{L}(\vec{x_1},...,\vec{x_N};\boldsymbol{\theta})\right] &=& \sum_{i=1}^{N}\ln\left[ \frac{f(\vec{x}_i;\boldsymbol{\theta})}{\int f({\vec{x}};\boldsymbol{\theta}) d{\vec{x}}}\right],\label{eq:typicalLL}
\end{eqnarray}
where the $\theta_i$ parameters are unknown constants to be inferred from the $S_e$ dataset.
The best estimation of $\boldsymbol{\theta}'$ parameters ($\boldsymbol{\hat{\theta}}$) are those maximizing the logarithmic likelihood ln$\mathcal{L}$ as
\begin{eqnarray}
    \frac{\partial}{\partial \theta_i}\ln \left[\mathcal{L}(\vec{x_1},...,\vec{x_N};\boldsymbol{\theta})\right]\bigg|_{\boldsymbol{\theta}=\boldsymbol{\hat{\theta}}}=0.
\end{eqnarray}

In practice if the experimental efficiency $\epsilon({\vec{x}})$ is not uniform, the data sample $S_e=\{\vec{x_1},...,\vec{x_N}\}$ follows the distribution of $f(\vec{x};\boldsymbol{\theta}')\epsilon({\vec{x}})$, then the likelihood can be constructed  as 

\begin{eqnarray}
    \ln \left[\mathcal{L}(\vec{x_1},...,\vec{x_n};\boldsymbol{\theta})\right] &=& \sum_{i=1}^{N}\ln\left[ \frac{f(\vec{x_i};\boldsymbol{\theta})\epsilon_i}{\int f(\vec{x};\boldsymbol{\theta})\epsilon({x}) d\vec{x}}\right],\label{eq:defaultLL}
\end{eqnarray}
where the per-event efficiency $\epsilon_i \equiv \epsilon(\vec{x}_i)$ is defined. The denominator $\int f(\vec{x};\boldsymbol{\theta})\epsilon({x}) d\vec{x}$ normalizes the distribution in the numerator. The efficiency function $\epsilon({\vec{x}})$ may be measured through a full Monte Carlo simulation of the experiment.
In this manuscript, another likelihood function is proposed which takes the inverse efficiency as the per-event weight
\begin{eqnarray}
    \ln \left[\mathcal{L}(\vec{x_1},...,\vec{x_n};\boldsymbol{\theta})\right] &=& \frac{\sum_{i=1}^{N}1/\epsilon_i}{\sum_{i=1}^{N}1/\epsilon_i^2} \sum_{i=1}^{N}\frac{1}{\epsilon_i}\ln\left[ \frac{f(\vec{x}_i;\boldsymbol{\theta})}{\int f(\vec{x};\boldsymbol{\theta}) d\vec{x}}\right].\label{eq:altLL}
\end{eqnarray}
The factor $C(\epsilon_i)\equiv\frac{\sum_i1/\epsilon_i}{\sum_i1/\epsilon_i^2}$ is used to correct the parameter uncertainties calculated using second derivatives at the maximum of the likelihood. 
According to the study in Ref.~\cite{Langenbruch:2019nwe} the correct factor $C(\epsilon_i)$ may require additional corrections, which can be studied using pesudoexperiments.
The advantage of the fit using the likelihood in Eq.~\ref{eq:altLL} is that the PDF normalization may be calculated analytically.
While the function in Eq.~\ref{eq:defaultLL} follows closely the definition of a likelihood fit, the likelihood in Eq.~\ref{eq:altLL} is also proved to be correct in the following in the likelihood principle.

\emph{Proof}:
The parameters $\boldsymbol{\hat{\theta}}$ that maximize the likelihood function meet the requirement
\begin{eqnarray}
    \frac{\partial}{\partial \theta_i}\int f(\vec{x};\boldsymbol{\theta}')\times \log \left[\frac{f(\vec{x};\boldsymbol{\theta})}{\int f(\vec{x}';\boldsymbol{\theta}) d\vec{x}'}\right] d\vec{x} 
    \Bigg|_{\boldsymbol{\theta}=\boldsymbol{\hat{\theta}}}=0,
\end{eqnarray}
which is equivalent to 
\begin{eqnarray}
 \frac{\partial}{\partial \theta_i}  \int \left[f(\vec{x};\boldsymbol{\theta}') \epsilon(\vec{x})\right]\frac{1}{\epsilon(\vec{x})}\log\left[ \frac{f(\vec{x};\boldsymbol{\theta})}{\int f(\vec{x}';\boldsymbol{\theta}) d\vec{x}'}\right] d\vec{x} \Bigg|_{\boldsymbol{\theta}=\boldsymbol{\hat{\theta}}}=0\label{eq:ideal-eff}
\end{eqnarray}
for any arbitrary efficiency function $\epsilon(\vec{x})$.
With a dataset $S_e\equiv\{\vec{x}_1,...,\vec{x}_N\}$ sampled according to the distribution $f(\vec{x};\boldsymbol{\theta'})\epsilon(\vec{x})$, Eq.~\ref{eq:ideal-eff} can be estimated approximated by
\begin{eqnarray}
 \frac{\partial}{\partial \theta_i}\sum_{\vec{x}_i\in S_e}    \frac{1}{\epsilon(\vec{x}_i)}\log\left[  \frac{f(\vec{x}_i;\boldsymbol{\theta})}{\int f(\vec{x}';\boldsymbol{\theta}) d\vec{x}'} \right]\Bigg|_{\boldsymbol{\theta}=\boldsymbol{\hat{\theta}}}=0\label{eq:dataset-eff}
\end{eqnarray}
at the limit of the $S_e$ sample size $N\to\infty$. The corresponding likelihood in Eq.~\ref{eq:dataset-eff} is equivalent to that of Eq.~\ref{eq:altLL}, apart from a factor to correct parameter uncertainties.  
This establishes that the likelihood function presented in Eq.~\ref{eq:altLL} is an estimator that aligns with the original definition of a maximum likelihood fitter.

\section{Pseudo-experiment studies}
\begin{figure}
    \centering
    \includegraphics[width=0.8\textwidth]{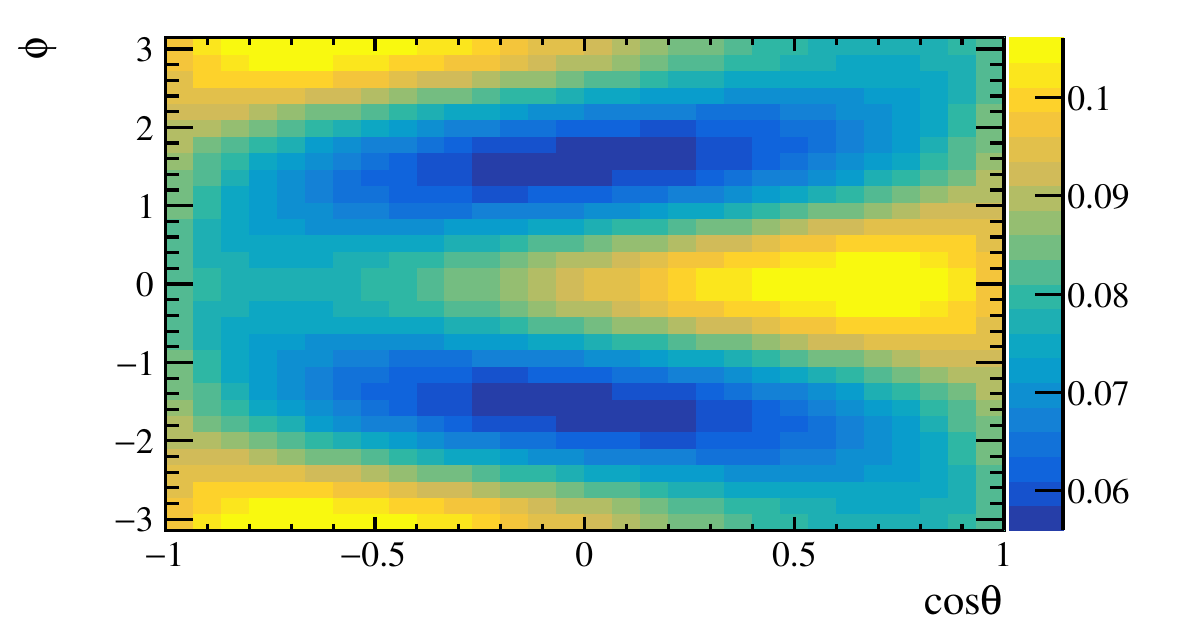}
    \includegraphics[width=0.8\textwidth]{figs/toys1eff.pdf}
    \includegraphics[width=0.8\textwidth]{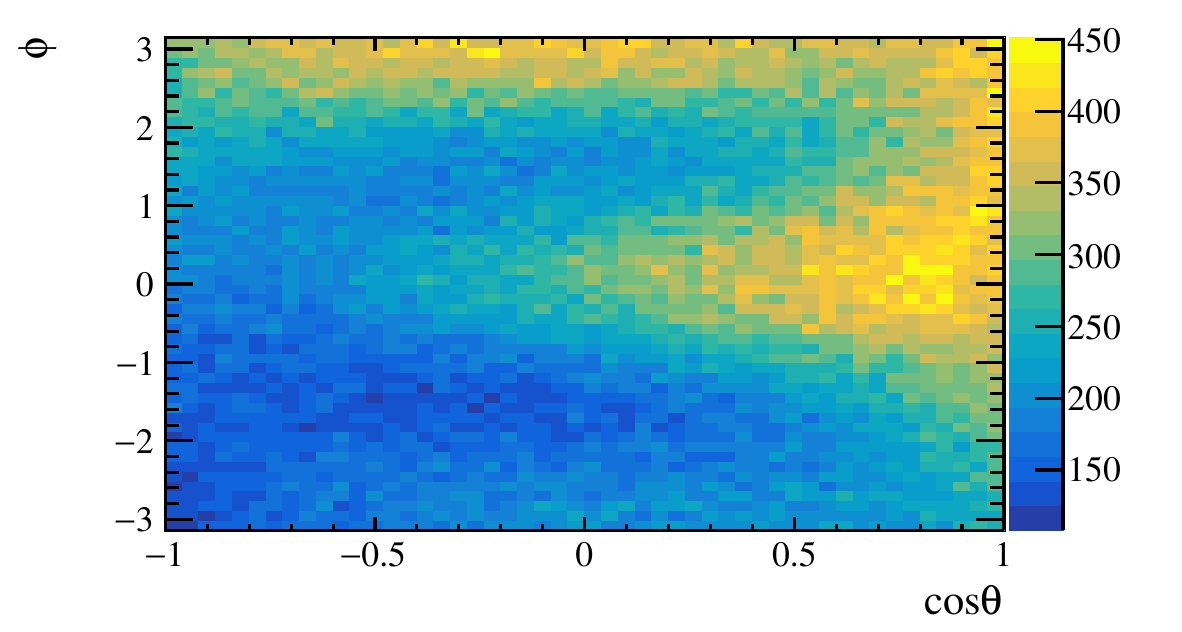}
    \caption{(Top) The true distribution of $f(\cos\theta,\phi|\lambda_\theta,\lambda_{\theta\phi},\lambda_\phi)$, (middle) the efficiency distribution $\epsilon(\cos\theta,\phi)$ and the distribution of pseudo-data sampled according to the product the true PDF and the efficiency distribution. Note the efficiency distribution is not normalized. (Bottom) The distribution, comprising approximately 700,000 pseudo-data points, is generated by multiplying the PDF distribution found in Equation \ref{eq:pdfpolar} with an efficiency distribution.}
    \label{fig:toy}
\end{figure}

Studies with pseudo-experiments are performed to verify the performance of the modified likelihood fitter in Eq.~\ref{eq:altLL}. The PDF describing heavy quarkonium polarization in the $\mu^+\mu^-$ final state~\cite{LHCb:2014brf},
\begin{eqnarray}
    f(\cos\theta, \phi|\lambda_\theta,\lambda_{\theta\phi},\lambda_\phi)=\frac{1+\lambda_\theta \cos^2\theta+\lambda_{\theta\phi}\sin2\theta \cos\phi+\lambda_\phi \sin^2\theta \cos2\phi}{4\pi(3+\lambda_\theta)/3},\label{eq:pdfpolar}\\\text{with }\cos\theta\in [-1,1],\phi\in(-\pi,\pi],\nonumber
\end{eqnarray}
is used to generate pseudo-data.
The following arbitrary polarization parameters are chosen, $\lambda_\theta=0.2, \lambda_\phi=0.25, \lambda_{\theta\phi}=0.2$. The two dimensional $(\cos\theta, \phi)$ distribution is shown in the top plot of Fig.~\ref{fig:toy}.  An arbitrary non-uniform distribution is used to represent experimental efficiency as shown in middle of Fig.~\ref{fig:toy} (unnormalized). The bottom plot in Fig.~\ref{fig:toy} correspond to the distribution of about 700 000 events sampled according to the PDF in Eq.~\ref{eq:pdfpolar} times the efficiency distribution.

\begin{figure}
    \centering
    \includegraphics[width=0.48\textwidth]{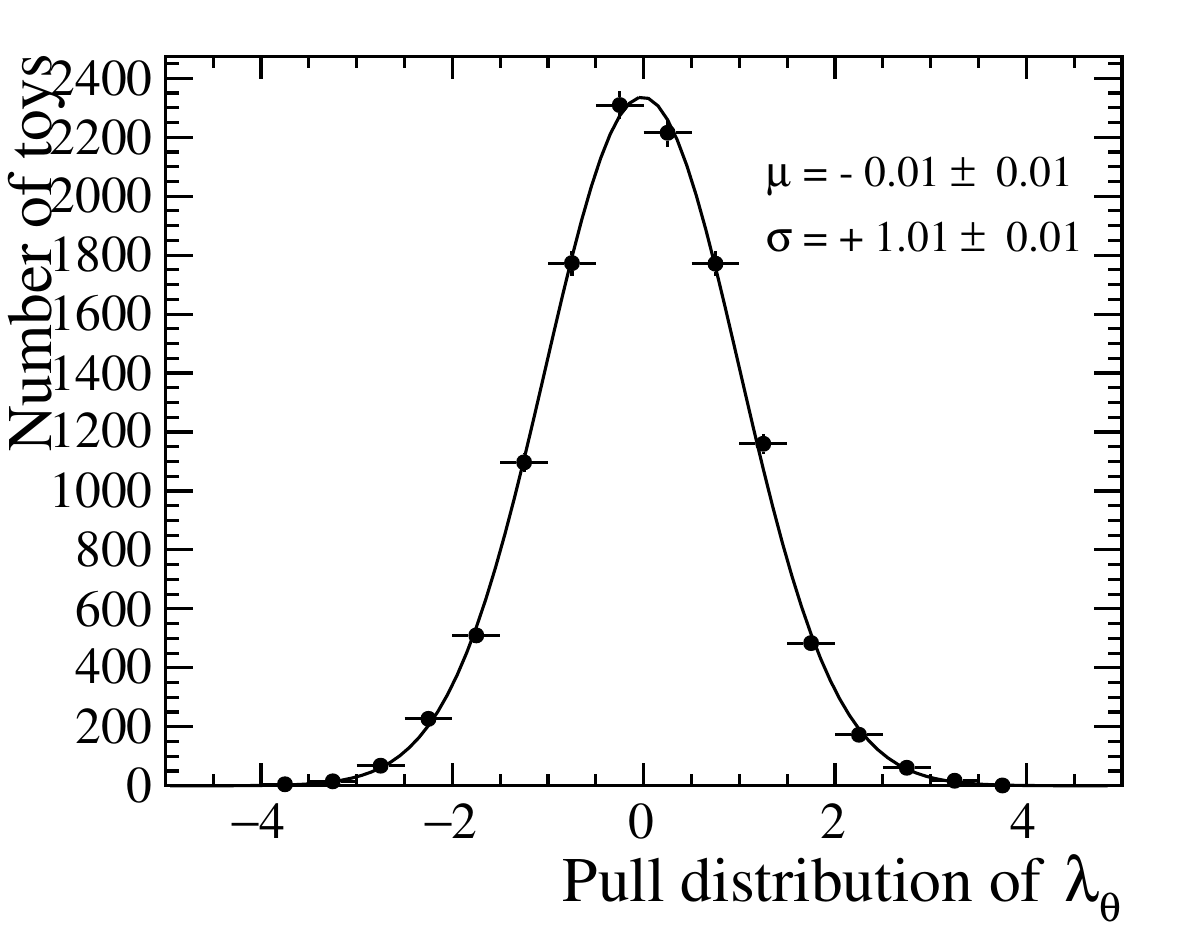}
    \includegraphics[width=0.48\textwidth]{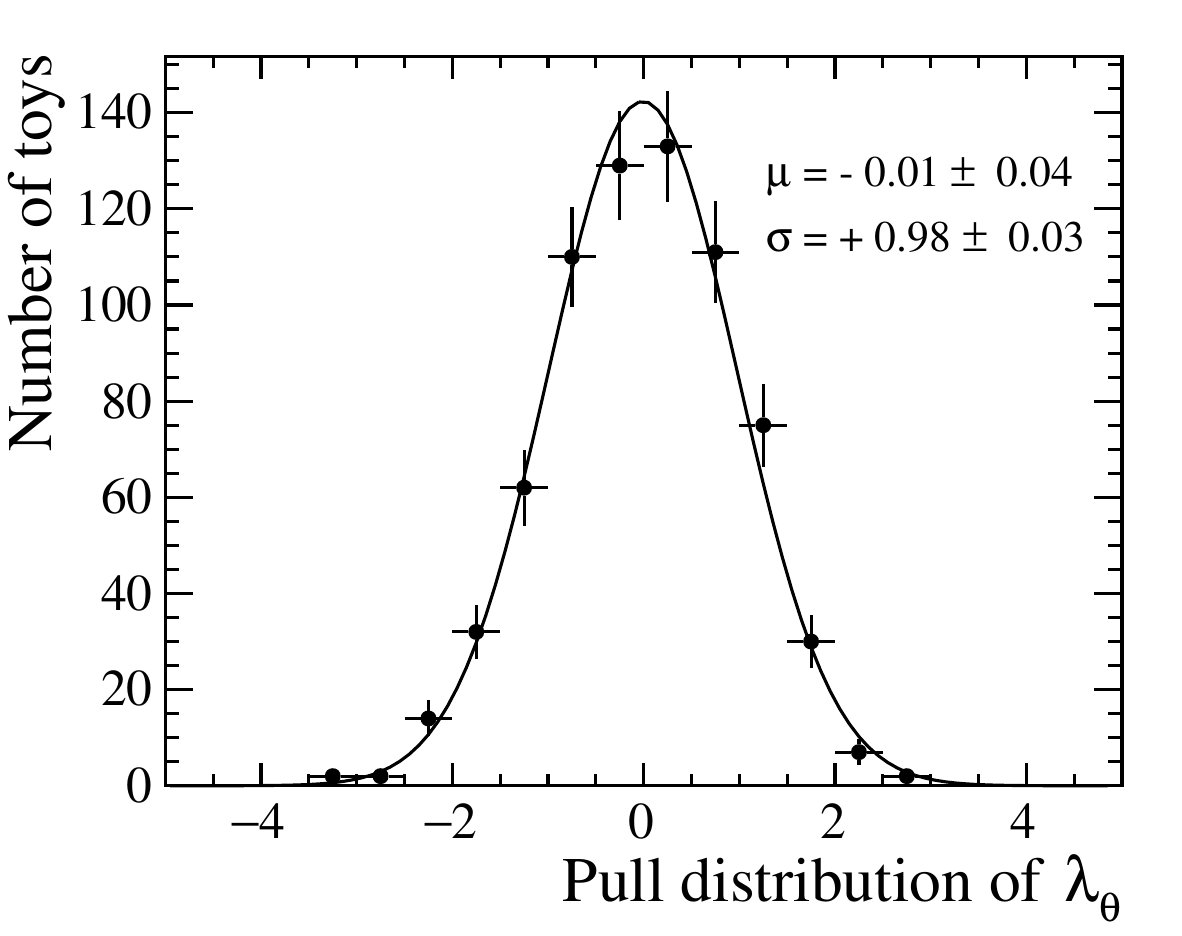}
    \includegraphics[width=0.48\textwidth]{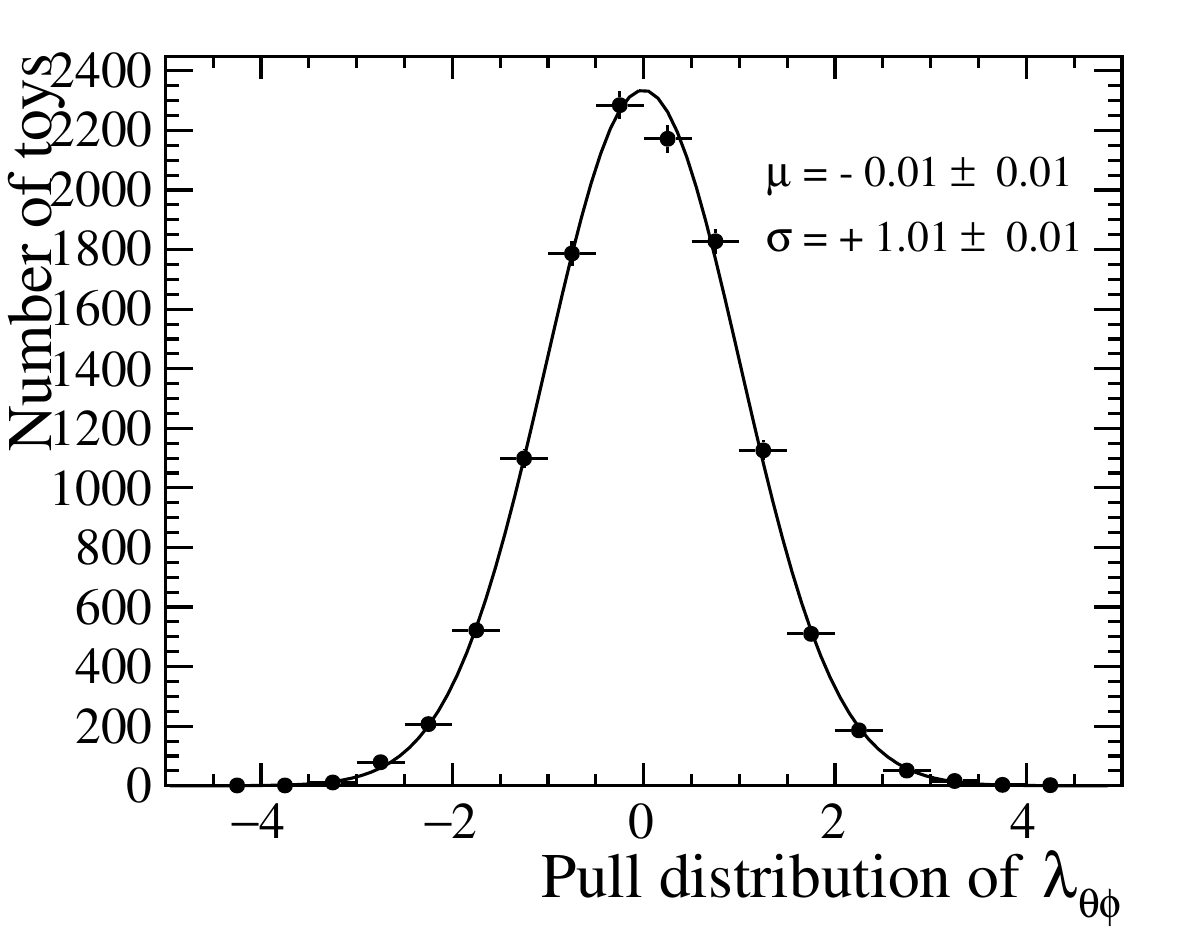}
    \includegraphics[width=0.48\textwidth]{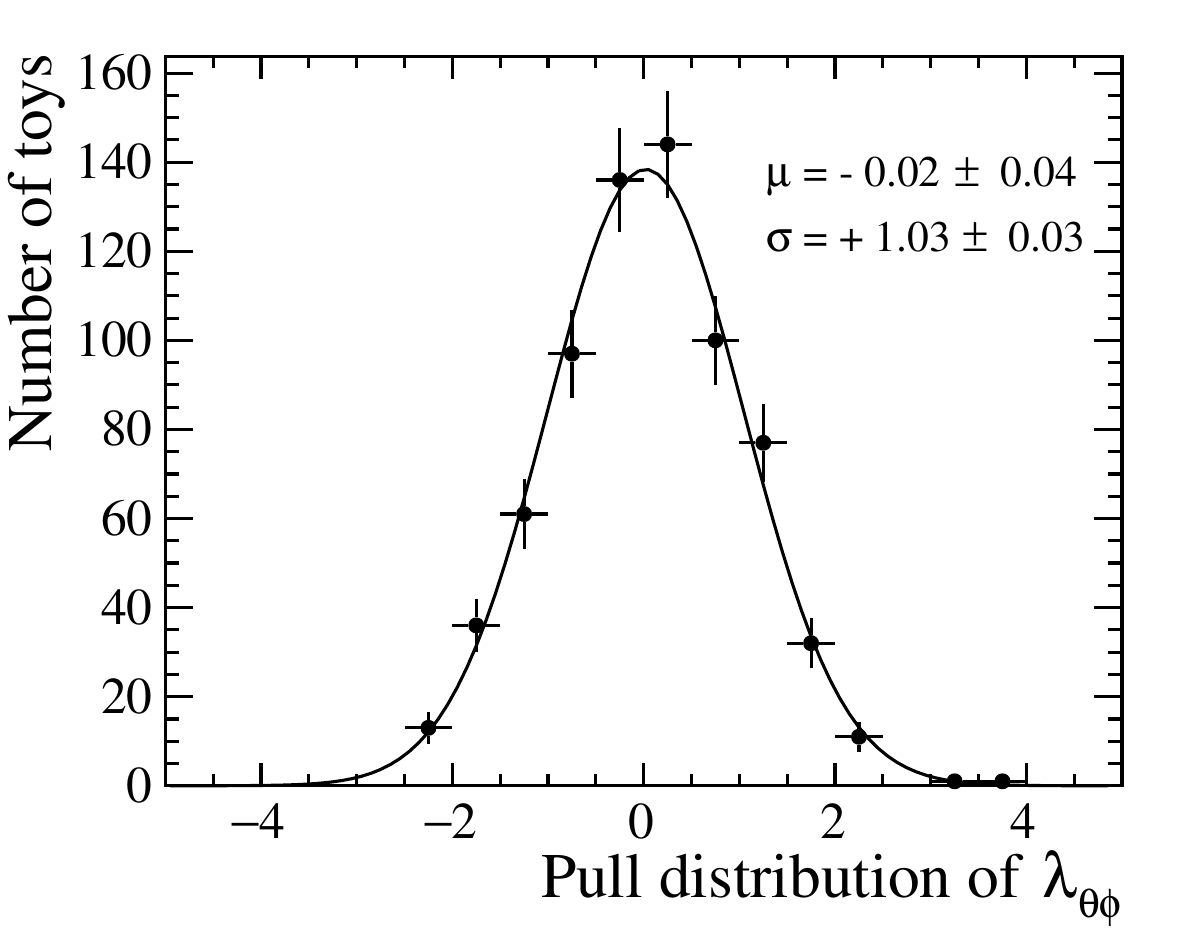}
    \includegraphics[width=0.48\textwidth]{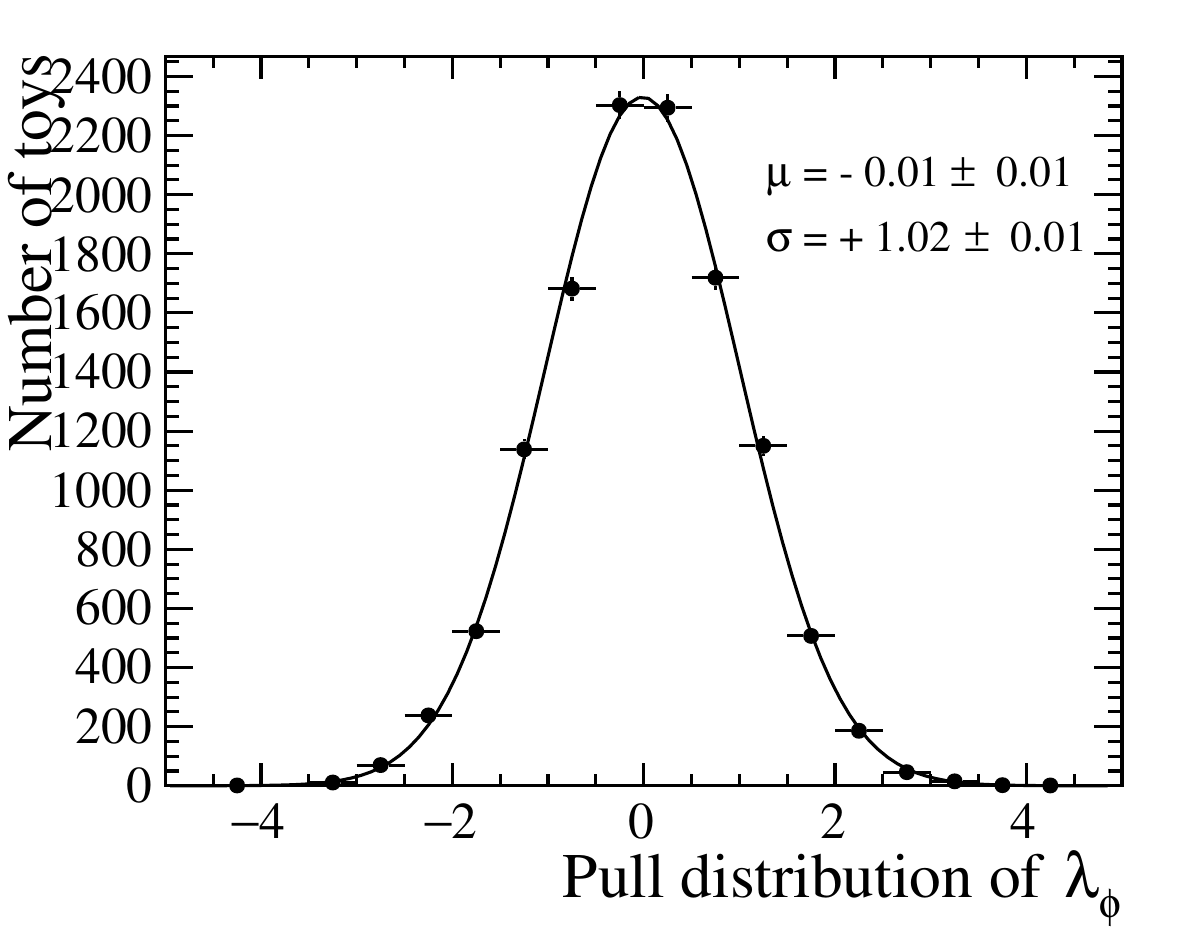}
    \includegraphics[width=0.48\textwidth]{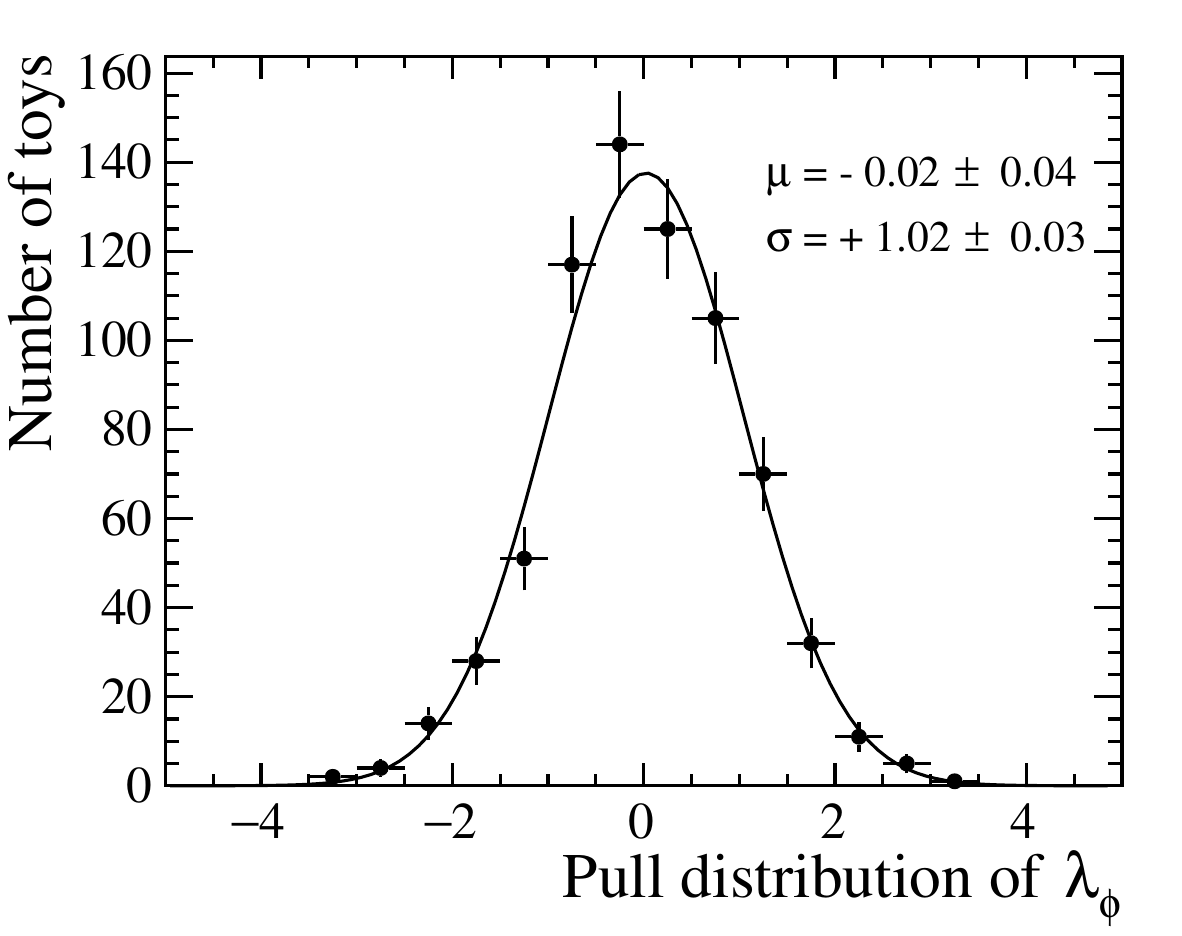}
    \caption{The distributions of parameter pulls for fits to pseudo-data. Left: Fits are performed with efficiencies considered as weights to the likelihood. Right: Fits are performed with the efficiency distribution included explicitly in the PDF. Top: the $\lambda_\theta$ parameter, Middle: the $\lambda_{\theta\phi}$ parameter, Bottom: the $\lambda_{\phi}$ parameter.}
    \label{fig:fit3toys}
\end{figure}

A large ensemble of pseudo-data samples are generated.
Each sample is fitted with the likelihood defined in Eq.~\ref{eq:defaultLL} and Eq.~\ref{eq:altLL} respectively, to obtain the $\lambda$ parameters and their uncertainties. From the result, a pull, defined as $\frac{\lambda-\lambda_\text{input}}{\delta_\lambda}$, can be calculated from each $\lambda$ parameter for each pseudo-data sample, where $\lambda\pm\delta\lambda$ is provided by  the maximum likelihood fit and $\lambda_\text{input}$ is the value used to generate the pseudo-data.  It is found that for each pseudo-data, the uncertainties of the $\lambda$ parameters are consistent between the two likelihood fitters.
The distribution of pulls for the pseudo-data ensemble is shown in
Fig.~\ref{fig:fit3toys}, comparing the two maximum likelihood fitters. Only 1000 toys are studied for the likelihood defined in Eq.~\ref{eq:defaultLL}, due to the computational power required by the numeric method employed to carry out the PDF normalization integral. For the efficiency-weighted likelihood fit,  studies with ten times more toys are carried out as no additional PDF normalization is needed with respect to Eq.~\ref{eq:pdfpolar}.

Each pull distribution is fitted with a Gaussian distribution, which turns out to be consistent with the standard normal distribution. It confirms that the maximum likelihood function defined in Eq.~\ref{eq:defaultLL} yield the correct parameter estimations and uncertainties. 

\begin{figure}
    \centering\includegraphics[width=0.3\textwidth]{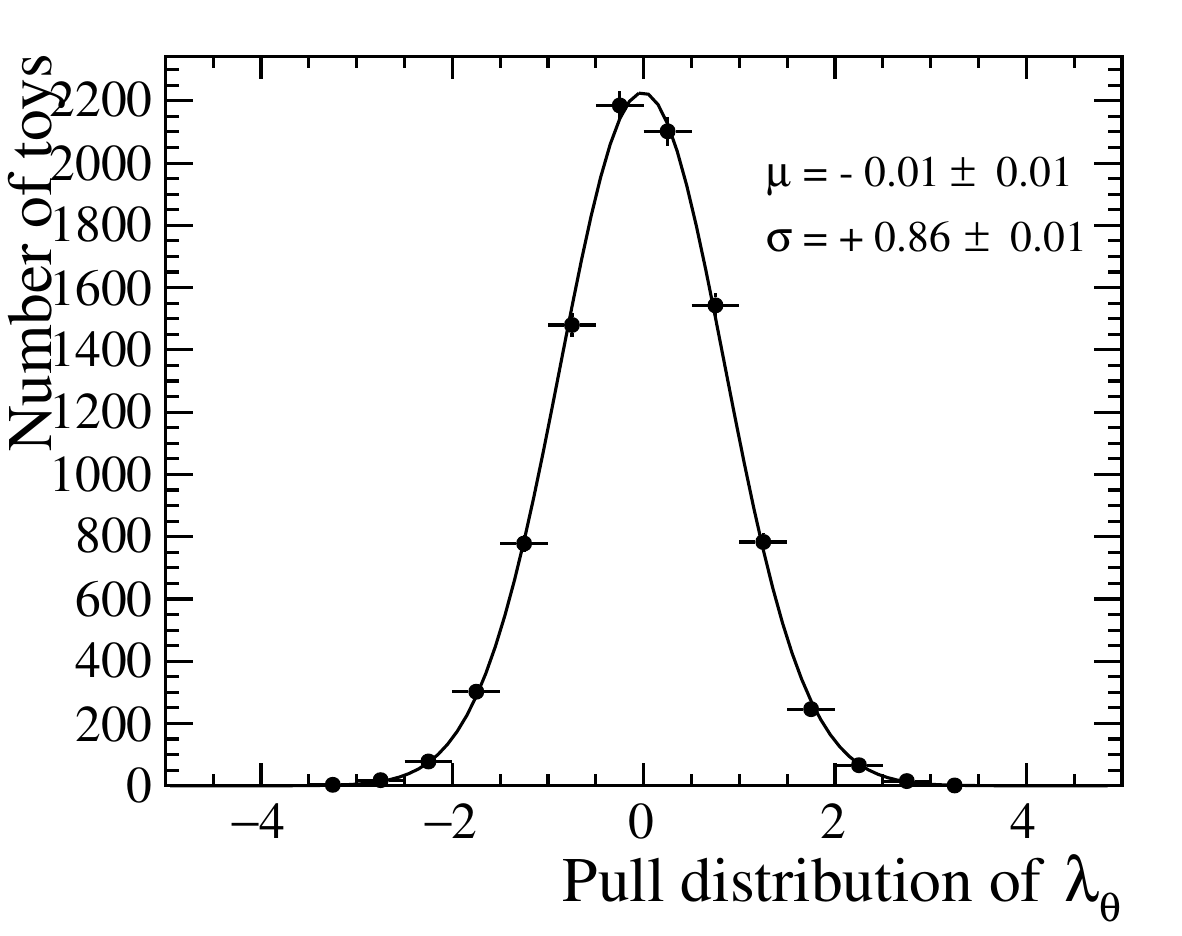}
    \includegraphics[width=0.3\textwidth]{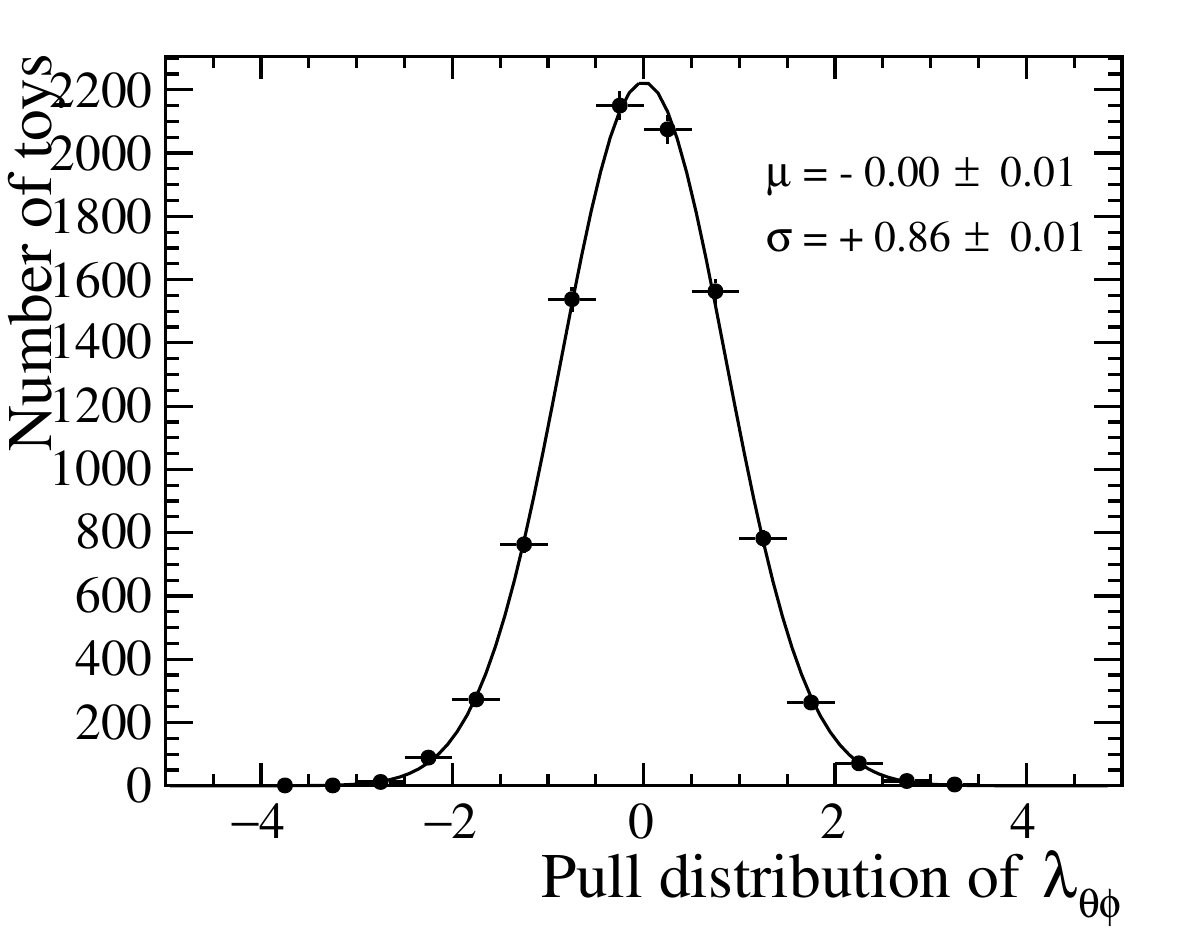}
    \includegraphics[width=0.3\textwidth]{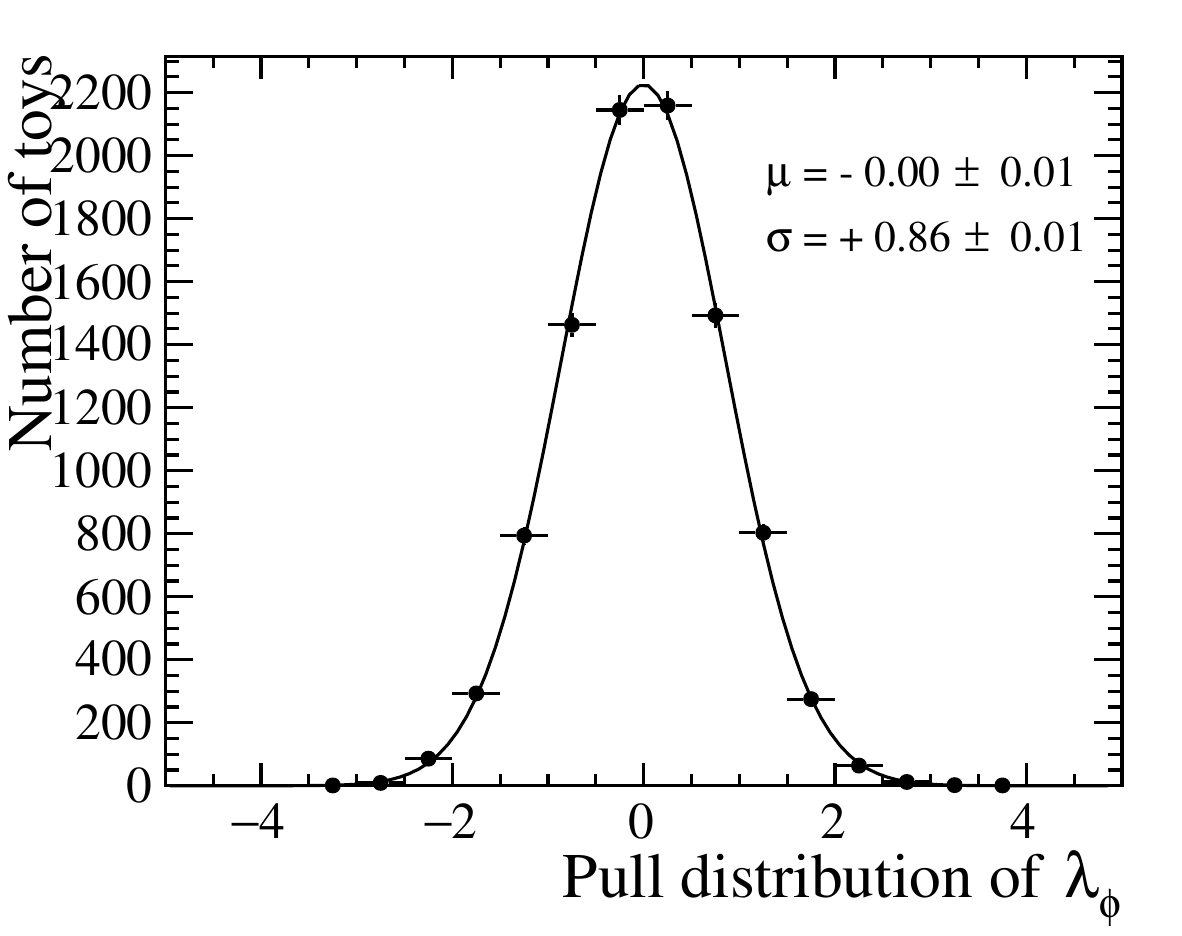}
    
    \caption{Parameter pull distributions for fits to pseudo-data without the term $C_{\epsilon}$ in Eq.~\ref{eq:altLL} to correct uncertainties.  Left: the $\lambda_\theta$ parameter, Middle: the $\lambda_{\theta\phi}$ parameter, Right: the $\lambda_{\phi}$ parameter.}
    \label{fig:fit4errs}
\end{figure}

To cross-check how well the global weight $C(\epsilon_i)\equiv\frac{\sum_i1/\epsilon_i}{\sum_i1/\epsilon_i^2}$ in Eq.~\ref{eq:altLL} correct parameter systematic uncertainties, separate fits are performed with this factor removed.
Fig.~\ref{fig:fit4errs} shows the pull distributions of the fit results. In comparison with the results in Fig. \ref{fig:fit3toys}, it is concluded that without the correction factor the parameter means are unbiased as expected, but significant biases are associated with their uncertainties. The pull distributions can be used make corrections to the fit results. In any case, it is suggested to validate the fits with pseudo-experiments.

\section{Summary}
This paper discusses a new method of maximum likelihood fit when experimental efficiencies are present for data. With the per-event efficiency as a weight to the likelihood, there is no need to include a further efficiency distribution in the PDF of the likelihood. A global factor formed by per-event efficiencies can be included to the likelihood function to correct the parameter uncertainties provided by the likelihood fitter. The new method has the advantage that, for specific models, the PDF normalization can be calculated analytically, thus reducing computing consumption.
Pseudo-experiment studies are performed, which suggest the efficiency-weighted likelihood results in unbiased estimations of parameters as well as their uncertainties. 
For more complex examples, it is suggested to always use pseudo-experiments to valid the fitter.

\clearpage

\addcontentsline{toc}{section}{References}

\bibliographystyle{mybst}
\bibliography{main}

\end{document}